\documentclass{article}
\usepackage[top=30truemm,bottom=30truemm,left=25truemm,right=25truemm]{geometry}
\usepackage{lscape} 
\usepackage{comment} 

\usepackage{multirow}
\usepackage{graphicx}
\usepackage{bm}
\usepackage{float}
\usepackage{amsmath}
\usepackage{amssymb}
\usepackage{amsthm}
\newtheorem{theo}{Theorem}

\begin{document}

\title{Bi-clustering for time-varying relational count data analysis}

\author{Satoshi Goto\thanks{Graduate School of Culture and Information Science, Doshisha University, Tataramiyakodani 1-3, Kyotanabe City, Kyoto, Japan.}\and
		Mariko Takagishi\footnotemark[1]  \and
			Hiroshi Yadohisa\thanks{Department of Culture and Information Science, Doshisha University, Tataramiyakodani 1-3, Kyotanabe City, Kyoto, Japan} 
	}

\date{}
\maketitle

%
\begin{abstract}
	Relational count data are often obtained from sources such as simultaneous purchase in online shops and social networking service information. Bi-clustering such relational count data reveals the latent structure of the relationship between objects such as household items or people. When relational count data observed at multiple time points are available, it is worthwhile incorporating the time structure into the bi-clustering result to understand how objects move between the cluster over time. In this paper, we propose two bi-clustering methods for analyzing time-varying relational count data. The first model, the dynamic Poisson infinite relational model (dPIRM), handles time-varying relational count data. In the second model, which we call the dynamic zero-inflated Poisson infinite relational model, we further extend the dPIRM so that it can handle zero-inflated data. Proposing both two models is important as zero-inflated data are often encountered, especially when the time intervals are short. In addition, by explicitly deriving the relevant full conditional distributions, we describe the features of the estimated parameters and, in turn, the relationship between the two models. We show the effectiveness of both models through a simulation study and a real data example.
	
\end{abstract}


\section{Introduction}
\label{intro}
Relational data contain information about the relationship between objects and may be obtained from, for example, point-of-sale (POS) information and the social networking service (SNS) community. These types of data are often represented as a matrix in which each element indicates the relationship between the row and column elements (e.g., the presence or absence of a relationship or its strength). Bi-clustering is one way to cluster elements along the row and column dimensions of a matrix simultaneously \cite{biclu1,biclu2}. By applying bi-clustering to relational data, we can ascertain the latent structure of the relationship. Indeed, many probabilistic models using bi-clustering of relational data have been proposed \cite{IRM,SBM,SIRM}.

Another situation commonly found in practice is when relationships change over time. For example, in the case of simultaneous purchases of products, the relationships between them may change according to events such as the season. In the SNS community, on the contrary, human relationships change through events such as career moves. In this way, the relationship between objects often changes through an event. In this paper, we call relational data obtained at multiple time points ``time-varying relational data,'' as in \cite{dIRM}.

A naive approach to bi-clustering time-varying relational data is to apply bi-clustering to the data at each time step separately. However, in practice, this leads to unstable clustering results \cite{BAFC}, and it becomes difficult to interpret clusters across time points since cluster structures are inconsistent over time. On the contrary, clustering using a time-varying structure \cite{JAT,DMMB} can help detect changes in relationships due to specific events.

Several bi-clustering methods exist for time-varying relational data \cite{BAFC,JAT,DMMB}. First, the dynamic infinite relational model (dIRM) \cite{dIRM}, which extends HMM-based method \cite{hmm1,hmm2} to relationship data, is especially useful since it can automatically determine the number of clusters from the data. The dIRM method can handle relational data containing either a zero or a one (essentially whether a relationship exists). However, in practice, we often obtain relational count data. For example, if each element of a data matrix represents how many times two products (corresponding to the row and column of the matrix) are simultaneously purchased, this data can be considered to be relational count data between the products. Since count data contain more information than binary data, it is preferable to use a model that can handle relational count data when we have count information instead of only zeros and ones.

Based on the foregoing, in this paper, we propose two new methods for the bi-clustering of count data under time changes. First, we extend the Poisson infinite relational model (PIRM) \cite{PIRM}, a bi-clustering method for relational count data that does not consider bi-clustering relationships that change over time, to be able to handle time changes to a model. We call this first model the dynamic Poisson infinite relational model (dPIRM). Second, we propose a dPIRM that can handle data containing many zeros, which we call the dynamic zero-inflated Poisson infinite relational model (dZIPIRM). The latter extension is especially important for count data, as zero-inflated count data are often obtained in practice, especially when the interval between observation times is short \cite{bayeszip,zi1,zi2}.

The remainder of the paper is organized as follows. In Section 2, we begin by considering a bi-clustering model of time series count data without zero-inflated characteristics. Next, we propose a bi-clustering model for zero-inflated time-varying count data. In Section 3, we derive the posterior distributions of the relevant parameters and clarify their properties. In Section 4, we illustrate how our proposed models can handle numerically simulated data. In Section 5, we show how the two proposed bi-clustering methods may be used with real data. Finally, Section 6 concludes.
\section{Bi-clustering for time-varying count data}
\label{sec:2}
In this section, we propose two bi-clustering methods. We first introduce one for time-varying count data and then the zero-inflated Poisson model (ZIP) \cite{ZIP} to extend the first method to be able to handle zero-inflated count data.
\subsection{The dPIRM}
\label{sec:2.1}
We denote the data by $X=\bigl\{x_{tij}\in \{0,1,2,\cdots\}; 1\leq i \leq N_1, 1\leq j \leq N_2,1\leq t \leq T\bigr\}$, where $N_1$ represents the number of row objects, $N_2$ represents the number of column objects, and $T$ represents the time horizon. We can assume that $N_1=N_2$ without loss of generality. In addition, since relational count data are not necessarily limited to an undirected graph, we allow asymmetric relations (i.e., $x_{tij} \neq x_{tji}$ for all $i,j \,( i\neq j)$).

We propose a bi-clustering model of time-varying count data under the following three assumptions:
\begin{description}
	\item[(A1)] Cluster structures remain consistent over time.
	\item[(A2)] Objects can move between clusters at each time step.
	\item[(A3)] The strength of the relationship between clusters can be expressed from zero to infinity.
\end{description}
We describe this dPIRM as follows:
\begin{align}
	\label{eq:stick}
	\bm{\beta}|\gamma &\sim \mathrm{Stick}(\gamma)\\
	\label{eq:DP}
	\bm{\pi}_{tk}|\alpha_0,\kappa,\bm{\beta} &\sim \mathrm{DP}\left(\alpha_0+\kappa,\frac{\alpha_0\bm{\beta}+\kappa\delta_k}{\alpha_0+\kappa}\right)\\
	\label{eq:Zti}
	z_{ti}|z_{(t-1)i},\Pi_t &\sim \mathrm{Multinomial}(\bm{\pi}_{tz_{(t-1)i}})\\
	\label{eq:dplamm}
	\lambda_{k\ell}|a,b &\sim \mathrm{Gamma}(a,b)\\
	\label{eq:dpxx}
	x_{tij}|Z_t,\{\lambda_{k\ell}\} &\sim \mathrm{Poisson}(\lambda_{z_{ti}z_{tj}}).
\end{align}
Here, $\Pi_t = \{\bm{\pi}_{tk} : k=1,2,\cdots\}$, $z_{ti}=k,k\in \{1,2,\cdots\}$.

In Equation (\ref{eq:stick}), the term ``Stick'' denotes a division formula known as the stick-breaking process \cite{stick}. This process can be described as $\beta_k=v_k\prod_{l=1}^{k-1}(1-v_l)$, where $v_{k} \sim \mathrm{Beta}(1,\gamma)$. $\gamma > 0$ is a shape parameter. $\bm{\beta}=(\beta_1,\beta_2,\cdots)$ represents time-averaged membership of the clusters (A1), and the $\bm{\beta}$ are guaranteed to satisfy $\sum_{k}\beta_k=1$. 

DP in Equation (\ref{eq:DP}) represents the Dirichlet process. $\alpha_0$ plays the role of a concentration parameter in the Dirichlet distribution. Thanks to the DP, the number of clusters is automatically determined.
$\pi_{tk\ell}$ in $\bm{\pi}_{tk}=(\pi_{tk1}$, $\pi_{tk2}$, $\cdots$, $\pi_{tk\ell},\cdots)$ represents the transition probability that an object belonging to cluster $k$ at time $t-1$ moves to cluster $\ell$ at time $t$.
$\pi_{tk\ell}$ is guaranteed to satisfy $\pi_{tk\ell}>0$ and $\sum_{\ell}\pi_{tk\ell}=1$. 
$\delta_k$ is an indicator function taking the value of one when it belongs to cluster $k$ and zero otherwise.
$\kappa>0$ adjusts how likely it is for objects to stay in the same cluster at each time step. In Equation (\ref{eq:Zti}), cluster assignment $Z_t=\{z_{ti}\}_{i=1}^{N}$ is generated with a multinomial distribution using $\bm{\pi}_{tk}$ to incorporate the time structure. This enables objects to move between clusters over time, following (A2).

In Equation (\ref{eq:dplamm}), $\lambda_{k\ell}$ represents the strength of the relationship between the cluster indicated by the $k$th row and $\ell$th column. $a$ and $b$ represent the shape and scale parameters of the Gamma distribution, respectively. By generating $\lambda_{k\ell}$ from a Gamma distribution, it is possible to represent relationships from zero to infinity, following (A3). In Equation (\ref{eq:dpxx}), count data $x_{tij}$ are generated with the parameter $\lambda_{z_{ti}z_{tj}}$.

Next, we consider the relationship between existing models (i.e., the dIRM and PIRM) and the dPIRM. For the dIRM, the Gamma distribution for the prior of $\lambda_{k\ell}$ in Equation (\ref{eq:dplamm}) of the dPIRM is replaced with a Beta distribution, while the Poisson distribution in Equation (\ref{eq:dpxx}) is replaced with the Bernoulli distribution. On the contrary, for the PIRM, the parameter $\bm{\beta}$ in Equation (\ref{eq:stick}) in the dPIRM is only used as a parameter for a multinomial distribution instead of $\bm{\pi}_{tz_{(t-1)i}}$ in Equation (\ref{eq:Zti}). Moreover, the PIRM framework does not consider the time structure as in Equation (\ref{eq:DP}).

Using the dPIRM, we can handle time-varying relational count data. However, in empirical applications, zero-inflated count data are often encountered, especially when the interval between time steps is short. In such cases, the approximation accuracy of the model is significantly low, as the dPIRM cannot handle zero-inflated count data. For this reason, we propose a model to address this issue in the next subsection.
\subsection{The dZIPIRM}
\label{sec:2.2}
In this section, we extend the dPIRM to handle zero-inflated count data by proposing the dZIPIRM. In addition to Assumptions (A1)--(A3), this extension further assumes the following:
\begin{description}
	\item[(A4)] Zero-inflated data are expressed by assuming mixture distributions of count data and zero data.
\end{description}
Given time-varying count data $x_{tij}\, (1\leq i,j \leq N, 1\leq t \leq T)$, the dZIPIRM model is defined as follows:
\begin{align}
	\label{d1}
	\bm{\beta}|\gamma &\sim \mathrm{Stick}(\gamma)\\
	\label{d2}
	\bm{\pi}_{tk}|\alpha_0,\kappa,\bm{\beta} &\sim \mathrm{DP}\left(\alpha_0+\kappa,\frac{\alpha_0\bm{\beta}+\kappa\delta_k}{\alpha_0+\kappa}\right)\\
	\label{d3}
	z_{ti}|z_{(t-1)i},\Pi_t &\sim \mathrm{Multinomial}(\bm{\pi}_{tz_{(t-1)i}})\\
	\label{wda}
	w_{tij}|c,d&\sim \mathrm{Beta}(c,d)\\
	\label{rdada}
	r_{tij}|w_{tij} &\sim \mathrm{Bernoulli}(w_{tij})\\
	\label{eq:dplam}
	\lambda_{k\ell}|a,b &\sim \mathrm{Gamma}(a,b)\\
	\label{eq:dpx}
	x_{tij}|Z_t,\{\lambda_{k\ell}\},r_{tij} &\sim \left\{ \begin{array}{ll}
		\mathrm{Degenerate}(x_{tij}=0) & (r_{tij}=0) \\
		\mathrm{Poisson}(\lambda_{z_{ti}z_{tj}}) & (r_{tij}=1)
	\end{array} \right ..
\end{align}
We assume that Equations (\ref{d1}), (\ref{d2}), (\ref{d3}), and (\ref{eq:dplam}) follow the same generator processes as the dPIRM. Equations (\ref{wda}), (\ref{rdada}), and (\ref{eq:dpx}) represent the extension of the dPIRM to zero-inflated count data. Before explaining the model in detail, we first describe how we handle zero-inflated data with this model.

First, we introduce the ZIP, which is often used to model zero-inflated data. In the Poisson distribution, we assume that zeros and nonzeros are obtained from the same generator process. On the contrary, in the ZIP, count data are generated from, mixture of a Poisson distribution with weight $w$ and a Degenerate distribution for $x=0$ with weight $1-w$. The Degenerate distribution for $x=0$ denotes a distribution of $P(x=0)=1$. Letting $x=x_{tij}$, $w=w_{tij}$, and $\lambda=\lambda_{z_{ti}z_{tj}}$, the probability function can be written as follows:
\begin{align}
	\label{eq:zero}
	P(x|\lambda,w)=(1-w)I(x=0)+w\frac{\lambda^xe^{-\lambda}}{x!}.
\end{align}
$I(\cdot)$ is an indicator function.

A naive approach to extending the dPIRM for zero-inflated data is to simply assume that data $x$ follows the ZIP. However, this assumption causes a problem when estimating the parameters. Specifically, to estimate the parameters using Gibbs sampling, a technique often used in Bayesian models, we need to derive the full conditional distribution. In general, however, it tends to be difficult to derive full conditional distributions explicitly when the probability is represented as a sum, as in Equation (\ref{eq:zero}) \cite{zipdif}. However, if we can derive the full conditional distribution, we can easily interpret the parameter properties. Therefore, to ensure tractability and interpretability, we introduce the latent variable $r$ \cite{ZIPhmm,bayzi}. Letting $r=r_{tij}$, we consider a model by setting the distribution as follows:
\begin{align}
	\label{eq:r}
	r|w &\sim \mathrm{Bernoulli}(w)\\
	\label{eq:xsitagau}
	x|\lambda,r&\sim \left\{ \begin{array}{ll}
		\mathrm{Degenerate}(x=0) & (r=0) \\
		\mathrm{Poisson}(\lambda) & (r=1)
	\end{array} \right ..
\end{align}
In Equation (\ref{eq:xsitagau}), the value of $r$ generated by Equation (\ref{eq:r}) determines whether the data follow the Poisson distribution or the Degenerate distribution for $x=0$. Thus, given $\lambda$ and $w$, the joint probability of $x$ and $r$ is
\begin{align}
	P(x,r|\lambda,w)&=P(x|\lambda,r)P(r|w)\notag\\
	&=\left(\frac{\lambda^xe^{-\lambda}}{x!}\right)^{r}I(x=0)^{1-r}w^{r}(1-w)^{1-r}\notag\\
	\label{wawotoru}
	&=\biggl\{w\frac{\lambda^xe^{-\lambda}}{x!}\biggr\}^{r}\biggl\{(1-w)I(x=0)\biggr\}^{1-r}.
\end{align}
In this way, the probability can be represented in the form of a product, and the joint probability of the parameters and data (i.e., $P(x,r,w,\lambda)$) is also expressed as a product. This solves the problem of the joint distribution in Equation (\ref{eq:zero}) being expressed as a sum.

Furthermore, when we sum out all the possible realized values of $r$ in Equation (\ref{wawotoru}), we obtain the formula on the left-hand side of the ZIP as follows:
\begin{align*}
	\sum_{r}P(x,r|\lambda,w)=(1-w)I(x=0)+w\frac{\lambda^xe^{-\lambda}}{x!}.
\end{align*}
The resulting formula is equivalent to the ZIP in Equation (\ref{eq:zero}). In other words, by assuming Equations (\ref{eq:r}) and (\ref{eq:xsitagau}), we obtain a model with the property of the ZIP while deriving an explicit posterior distribution. For this reason, the distributions in Equations (\ref{eq:r}) and (\ref{eq:xsitagau}) are assumed to hold in Equations (\ref{rdada}) and (\ref{eq:dpx}).

Since Equations (\ref{d1}), (\ref{d2}), (\ref{d3}), and (\ref{eq:dplam}) are equivalent to the dPIRM, we explain Equations (\ref{wda}), (\ref{rdada}), and (\ref{eq:dpx}) in detail. In Equation (\ref{wda}), we use the Beta distribution for the prior distribution of $w_{tij}$. Since the Beta distribution is the conjugate prior distribution of the Bernoulli distribution, we can analytically derive the posterior distribution, with $c>0$ and $d>0$ as the shape parameters. In Equations (\ref{rdada}) and (\ref{eq:dpx}), $r_{tij}$ determines whether $x_{tij}$ follows the Degenerate distribution for $x_{tij}=0$ or the Poisson distribution, following (A4).
This indicates that in this model, zero data, which is generated from the Degenerate distribution ($r_{tij} = 0$), is not taken into account for estimation.
In other words, when estimating the parameters in the dZIPIRM, we can remove the effect of zeros that follow the Degenerate distribution.
This intuitive interpretation will be justified by Theorem 1 in Section 3.2.
\begin{figure}[htbp]
	\begin{center}	
		\includegraphics[width=4.5in]{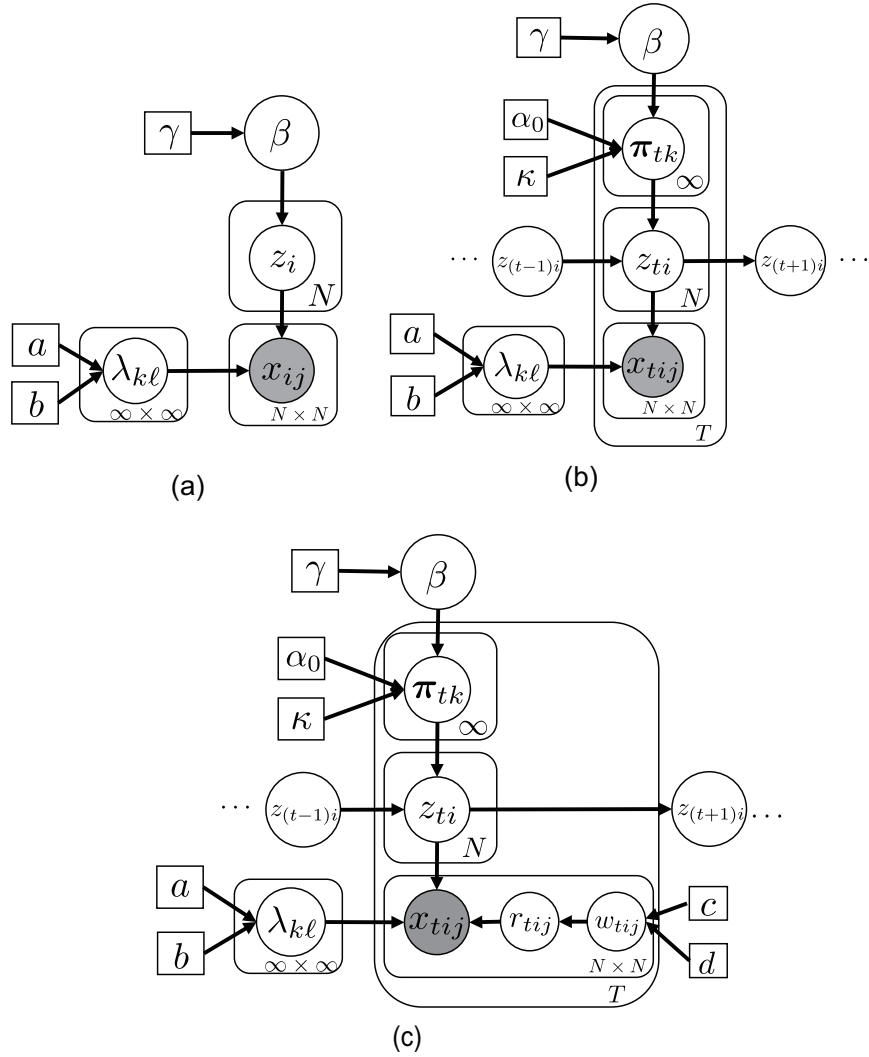}
		\caption{Graphical models of (a)PIRM, (b)dPIRM, (c)dZIPIRM.}
		\label{fig:gra}
	\end{center}
\end{figure}

Figure \ref{fig:gra} illustrates the parameter relations in the PIRM, dPIRM, and dZIPIRM. The circle nodes denote the variables, square nodes are the hyperparameters, and shaded nodes indicate the observations. The figure illustrates the differences between the data generation processes considered in each model, showing that the dZIPIRM has a structure in which $x_{tij}$ depends on $r_{tij}$.
\section{Sampling parameters}
\label{sec:3}
In this section, we derive the full conditional distribution of the parameter $\lambda_{k\ell}$ of the dPIRM as well as the parameters $\lambda_{k\ell}$, $w_{tij}$, and $r_{tij}$ of the dZIPIRM. We then explain their properties. Deriving the full conditional distributions allows us to understand the properties of the parameters when the other parameters and data are fixed. Moreover, by comparing $\lambda_{k\ell}$ for the dZIPIRM and dPIRM, we can appreciate how the models differ.

We use beam sampling \cite{beam,dIRM} to sample the parameters, as it is faster and more efficient than Gibbs sampling when there is strong series correlation between $z_{ti}$ and $z_{(t-1)i}$. Indeed, convergence is significantly slowed in this case for Gibbs sampling.
\subsection{Derivation of the full conditional distribution}
\label{sec:3.1}
In this section, we derive the full conditional distribution of the parameters $\lambda_{k\ell}$ in the dPIRM and $\lambda_{k\ell}$, $w_{tij}$, and $r_{tij}$ in the dZIPIRM. Since the full conditional distribution of the other parameters in the dPIRM and dZIPIRM are the same as in the dIRM, we refer the reader to \cite{dIRM} for the derivation of those distributions. In the derivation of the full conditional distribution, it is sufficient to consider the proportionality relation between the parameters for the calculation. To simplify the notation, we define the following set, $C_{k\ell}= \{t,i,j;z_{ti}=k,z_{tj}=\ell\}$.

First, we explain the parameters of the dPIRM. $\lambda_{k\ell}$ in the dPIRM is represented by the following proportionality relation:
\begin{align*}
	h(\lambda|x,Z) &\propto \lambda_{k\ell}^{a-1}e^{-b\lambda_{k\ell}}\prod_{t}\prod_{i;z_{ti}=k}\prod_{j;z_{tj}=\ell}e^{-\lambda_{z_{ti}z_{tj}}}\lambda_{z_{ti}z_{tj}}^{x_{tij}}\\
	&\propto \lambda_{k\ell}^{(a+\sum_{t,i,j\in C_{k\ell}}x_{tij})-1}e^{-[b+\sum_{t,i,j}I\{t,i,j\in C_{k\ell}\}]\lambda_{k\ell}}.
\end{align*}
Therefore, the full conditional distribution of $\lambda_{k\ell}$ in the dPIRM becomes
\begin{align}
	\label{dpirmlam}
	\lambda_{k\ell} &\sim \mathrm{Gamma}\left(a+\sum_{t,i,j\in C_{k\ell}}x_{tij},\,b+\sum_{tij}I\{t,i,j\in C_{k\ell}\}\right).
\end{align}

Next, we derive the full conditional distributions of $\lambda_{k\ell}$, $w_{tij}$, and $r_{tij}$ in the dZIPIRM. Unlike the dPIRM, the derivation of these full conditional distributions is not straightforward. However, thanks to the formulation of the dZIPIRM in Section \ref{sec:2}, setting conjugate prior distributions for the prior distribution of the parameters allows us to analytically derive the full conditional distribution of the parameters in the dZIPIRM. First, $\lambda_{k\ell}$ in the dZIPIRM is represented by the following proportionality relation:
\begin{align*}
	h(\lambda|w,r,x,Z) &\propto \lambda_{k\ell}^{a-1}e^{-b\lambda_{k\ell}}\prod_{t}\prod_{i;z_{ti}=k}\prod_{j;z_{tj}=\ell}e^{-\lambda_{z_{ti}z_{tj}}r_{tij}}\lambda_{z_{ti}z_{tj}}^{r_{tij}x_{tij}}\\
	&\propto \lambda_{k\ell}^{(a+\sum_{t,i,j\in C_{k\ell}}r_{tij}x_{tij})-1}e^{-[b+\sum_{t,i,j\in C_{k\ell}}r_{tij}]\lambda_{k\ell}}.
\end{align*}
Following a similar reasoning as in Equation (\ref{dpirmlam}), the full conditional distribution of $\lambda_{k\ell}$ in the dZIPIRM may be expressed as follows:
\begin{align}
	\label{zigolamne}
	\lambda_{k\ell} &\sim \mathrm{Gamma}\left(a+\sum_{t,i,j\in C_{k\ell}}r_{tij}x_{tij},\,b+\sum_{t,i,j\in C_{k\ell}}r_{tij}\right).
\end{align}
This expression may be further simplified as follows. From Equation (\ref{eq:dpx}):
\begin{align*}
	\begin{split}
		x_{tij}=0&\Longrightarrow
		\left\{
		\begin{array}{ll}
			r_{tij}=1 &(\mathrm{Poisson})\\
			r_{tij}=0&(\mathrm{Degenerate})
		\end{array}
		\right.,\\
		x_{tij} >0 &\Longrightarrow \quad r_{tij}=1 \,(\mathrm{Poisson}).
	\end{split}
\end{align*}
Using this, we obtain
\begin{align}
	\label{rtox}
	\sum_{t,i,j\in C_{k\ell}}r_{tij}x_{tij}=\sum_{t,i,j\in C_{k\ell}}x_{tij},
\end{align}
which we prove in Appendix. Using Equation (\ref{rtox}), Equation (\ref{zigolamne}) may be rewritten as follows:
\begin{align}
	\label{zigolam}
	\lambda_{k\ell} &\sim \mathrm{Gamma}\left(a+\sum_{t,i,j\in C_{k\ell}}x_{tij},\,b+\sum_{t,i,j\in C_{k\ell}}r_{tij}\right).
\end{align}
This expression is useful to compare the properties of the dPIRM and dZIPIRM, as explained in Section \ref{sec:3.2}.

Next, for the parameter $w_{tij}$ of the dZIPIRM, we use the proportionality relation:
\begin{align*}
	h(w|\lambda,r,x) &\propto \prod_{t}\prod_{i}\prod_{j}w_{tij}^{r_{tij}}(1-w_{tij})^{1-r_{tij}}w_{tij}^{c-1}(1-w_{tij})^{d-1}\\
	&\propto \prod_{t}\prod_{i}w_{tij}^{c+r_{tij}-1}(1-w_{tij})^{d+1-r_{tij}-1}.
\end{align*}
Therefore, the full conditional distribution of $w_{tij}$ becomes the Beta distribution as follows:
\begin{align}
	\label{zigow}
	w_{tij}&\sim \mathrm{Beta}\left(c+r_{tij},\,d+(1-r_{tij})\right).
\end{align}

Finally, we consider the full conditional distribution of the latent variable $r_{tij}$, for which we consider the proportionality relation:
\begin{align*}
	h(r|x,\lambda,w)&\propto
	\prod_{t}\prod_{i}\prod_{j}(e^{-\lambda_{z_{ti}z_{tj}}}(\lambda_{z_{ti}z_{tj}})^{x_{tij}})^{r_{tij}}\\
	&\quad\quad\quad\quad\times(I(x_{tij}=0))^{(1-r_{tij})}w_{tij}^{r_{tij}}(1-w_{tij})^{1-r_{tij}}\\
	&\propto\prod_{t}\prod_{i}\prod_{j}\{w_{tij}e^{-\lambda_{z_{ti}z_{tj}}}(\lambda_{z_{ti}z_{tj}})^{x_{tij}}\}^{r_{tij}}\\
	&\quad\quad\quad\quad\times\{(1-w_{tij})I(x_{tij}=0)\}^{(1-r_{tij})}.
\end{align*}
The full conditional distribution of $r_{tij}$ becomes
\begin{align*}
	r_{tij}&\sim \mathrm{Bernoulli}\left(\frac{w_{tij}e^{-\lambda_{z_{ti}z_{tj}}}(\lambda_{z_{ti}z_{tj}})^{x_{tij}}}{w_{tij}e^{-\lambda_{z_{ti}z_{tj}}}(\lambda_{z_{ti}z_{tj}})^{x_{tij}}+(1-w_{tij})I(x_{tij}=0)}\right).
\end{align*}

\subsection{Interpreting the properties of the model using the full conditional distribution}
\label{sec:3.2}
In this section, we clarify and compare the properties of the two models by studying the full conditional distributions of the parameters derived in Section \ref{sec:3.1}. First, we compare $\lambda_{k\ell}$ in the dPIRM and dZIPIRM to explain the properties of each model. The expected value and variance of $\lambda_{k\ell}$ in the dPIRM from Equation (\ref{dpirmlam}), denoted by $\lambda_{k\ell}^{(dP)}$, is expressed as follows:
\begin{align}
	\begin{split}
	\label{elam1}
	E\left[\lambda_{k\ell}^{(dP)}\right]&=\frac{a+\sum_{t,i,j\in C_{k\ell}}x_{tij}}{b+\sum_{t,i,j}I\{t,i,j\in C_{k\ell}\}},\\
	V\left[\lambda_{k\ell}^{(dP)}\right]&=\frac{a+\sum_{t,i,j\in C_{k\ell}}x_{tij}}{(b+\sum_{t,i,j}I\{t,i,j\in C_{k\ell}\})^2}.
	\end{split}
\end{align}
$\sum_{t,i,j\in C_{k\ell}}x_{tij}$ and $\sum_{t,i,j}I\{t,i,j\in C_{k\ell}\}$ represent the total value and number of $x_{tij}$ belonging to the $k$th row and $\ell$th column cluster, respectively. 
Ignoring $a$ and $b$, we can interpret Equation (\ref{elam1}) as indicating that the expected value of $\lambda_{k\ell}$ is the average value of $x_{tij}$ belonging to the $k$th row and $\ell$th column cluster through time. Note that $a$ and $b$ are hyperparameters that adjust the average value and must be determined in advance.
Further, from Equation (\ref{elam1}), we can see how $a$ and $b$ affect the estimated $\lambda_{k\ell}$. Specifically, $\lambda_{k\ell}$ becomes larger when $a>b$, while $\lambda_{k\ell}$ becomes smaller when $a<b$. In addition, the larger the values of $a$ and $b$, the larger the amount of prior information.
Thus, by deriving explicit relationship between parameters and hyperparameters, users can have intuition for how hyperparameters affect the estimation, and then it helps them select hyperparameters.

Next, we consider $\lambda_{k\ell}$ in Equation (\ref{zigolam}) of the dZIPIRM, denoted by $\lambda_{k\ell}^{(dZIP)}$. As before, the expected value and variance of $\lambda_{k\ell}^{(dZIP)}$ are derived to be
\begin{align}
\begin{split}
	\label{elam3}
	E\left[\lambda_{k\ell}^{(dZIP)}\right]&=\frac{a+\sum_{t,i,j\in C_{k\ell}}x_{tij}}{b+\sum_{t,i,j\in C_{k\ell}}r_{tij}},\\
	V\left[\lambda_{k\ell}^{(dZIP)}\right]&=\frac{a+\sum_{t,i,j\in C_{k\ell}}x_{tij}}{(b+\sum_{t,i,j\in C_{k\ell}}r_{tij})^2}.
	\end{split}
\end{align}
The numerator in Equation (\ref{elam3}) is the same as in Equation (\ref{elam1}). $\sum_{t,i,j\in C_{k\ell}}r_{tij}$ represents the number of $x_{tij}$ generated from the Poisson distribution ($r_{tij}=1$), which belongs to the $k$th row and $\ell$th column cluster. Ignoring $a$ and $b$, 
Equation (\ref{elam3}) indicates the average value of $x_{tij}$ considered to be generated from the Poisson distribution ($r_{tij}=1$), which belongs to the $k$th row and $\ell$th column cluster over time.

From Equations (\ref{elam1}) and (\ref{elam3}), when we compare $\lambda_{k,\ell}$ in the dPIRM and dZIPIRM, the following theorem is derived.
\begin{theo}
	\label{theotheo}
	\begin{align}
		\label{the1}
		E\left[\lambda_{k\ell}^{(dP)}\right]&=\alpha_{k\ell}E\left[\lambda_{k\ell}^{(dZIP)}\right],\\
		\label{the2}
		V\left[\lambda_{k\ell}^{(dP)}\right]&=\alpha_{k\ell}^2V\left[\lambda_{k\ell}^{(dZIP)}\right],
	\end{align}
	where
	\begin{align*}
		\alpha_{k\ell}=\frac{b+\sum_{t,i,j\in C_{k\ell}}r_{tij}}{b+\sum_{t,i,j}I\{t,i,j\in C_{k\ell}\}},
	\end{align*}
	and then $\alpha_{k\ell}\in [0,1]$.
\end{theo}
The proof of Theorem \ref{theotheo} is presented in Appendix.

From Equation (\ref{the1}) we obtain an intuitive interpretation of $\lambda_{k\ell}$. Ignoring $b$, $\alpha_{k\ell}$ represents the proportion of data that belongs to the $k$th row and $\ell$th column cluster and that are generated from a Poisson distribution ($r_{tij}=1$). That is, in the dZIPIRM, the more data in a cluster generated from the Degenerate distribution, the larger $\lambda_{k\ell}^{(dZIP)}$ tends to be compared with $\lambda_{k\ell}^{(dP)}$. Conversely, the more data generated from the Poisson distribution, the closer $\lambda_{k\ell}^{(dZIP)}$ tends to be to $\lambda_{k\ell}^{(dP)}$. In short, under the estimation framework using the full conditional distribution, the estimated $\lambda_{k\ell}^{(dZIP)}$ tends to become larger than $\lambda_{k\ell}^{(dP)}$ because in the dZIPIRM, the zero data, which are considered to be generated from the Degenerate distribution, are not taken into account to estimate $\lambda_{k\ell}^{(dZIP)}$. This theoretical result is consistent with the model interpretation in Section 2.

From Equation (\ref{the2}), we may also say that the dispersion of $\lambda_{k\ell}^{(dZIP)}$ tends to be larger than that of $\lambda_{k\ell}^{(dP)}$. However, as shown in the simulation study in Section 4, this result does not necessarily affect the clustering result.

Next, we consider $w_{tij}$ in Equation (\ref{zigow}). Again, the expected value and variance of $w_{tij}$ is expressed as
\begin{align}
	\label{w1}
	\begin{split}
		E[w_{tij}]&=\frac{c+r_{tij}}{c+d+1},\\
		V[w_{tij}]&=\frac{(c+r_{tij})(d+(1-r_{tij}))}{(c+d+1)^2(c+d+2)}.
	\end{split}
\end{align}
From Equation (\ref{w1}), we can see how the hyperparameters $c$ and $d$ affect the estimation. Especially, if $c$ and $d$ are set to be equal and small, the estimation of $w_{tij}$ is less affected by the hyperparameters $c$ and $d$.
\section{Simulation Study}
\label{sec:4}
In this section, we conduct a simulation to evaluate the fit to the data and clustering accuracy of the dPIRM and dZIPIRM.
\subsection{Data generation and setting}
\begin{figure}[htbp]
	\begin{center}	
		\includegraphics[width=3in]{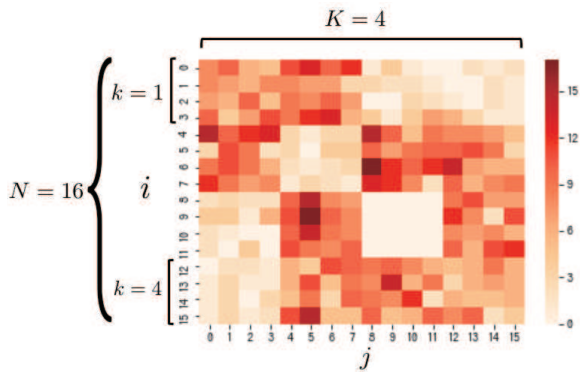}
		\caption{Example of the synthetic datasets ($t=1$)}
		\label{fig:simu}
	\end{center}
\end{figure}

We prepared several synthetic datasets. To create these datasets, we first stipulated that the number of clusters is $K=4$, the number of time steps is $T=5$, and the number of objects is $N=16$. Next, we generated $\lambda_{k\ell}$ from the discrete uniform distribution $\mathrm{Uniform}\{0,9\}$ ($t=1,\cdots,T$, $k,\ell=1,\cdots,K$), and then $x_{tij}$ from $\mathrm{Poisson}(\lambda_{z_{ti}z_{tj}})$ ($i,j = 1,\cdots,N$, $z_{ti}=1,\cdots,K$), as shown in Figure \ref{fig:simu}. When the time changes from $t$ to $t+1$, we randomly move objects to other clusters in three patterns of movement ratio $m$, where $m = 0.1, 0.2, 0.3$. Finally, we randomly convert the numerical value of the data into zero with a zero ratio $s$, where $s = 0.3, 0.5, 0.7$. We thus generate data for all nine ($=3\times 3$) settings.

To estimate the parameters, we need to prespecify the hyperparameters. In Section \ref{sec:3.2}, we showed the effect of the hyperparameters on the expected value. In this simulation, since there is no prior information, we reduce the influence of the hyperparameters as much as possible. Therefore, to reduce the influence of hyperparameters, we set $a=b=1$. Similarly, we set $c=d=1$ as well as $\kappa=\gamma=\alpha_0=1$.

We compare the PIRM, dPIRM, and dZIPIRM in this scenario by sampling the parameters 300 times for each method. We use two measurements for the evaluation. One is the Rand index (RI), which computes the similarity between the true and estimated clustering results \cite{RI,dIRM}. The RI is coded one (zero) when two clustering results match (do not match). As with the computation of the RI in the dIRM, we compute the RI between the assignment of correct answers $Z_{t}$ and the estimated $\hat{Z}_t$ for each time step, and then average the RI for $T$ time points. 
For the evaluation of cluster allocation in PIRM, where one cluster assignment is obtained in each time step (and thus T cluster assignments in total), we calculate $T$ RIs between the cluster assignment obtained by PIRM and true cluster assignment $Z_t$ at each time point, and pick up the highest RI result of all $T$ RI results.
The second measurement is log-likelihood, for which a large value indicates that the model fits the data well \cite{dIRM}.

We calculate the mean and standard deviation for the two measurements above for the data generated for 50 data points in each of the nine settings described earlier.
\subsection{Results}
\renewcommand{\arraystretch}{1.5}
\begin{table}[]
	\caption{RI results}
	\begin{center}
		\begin{tabular}{cc|lll}\hline
			\multicolumn{1}{l}{$m$} & $s$ & PIRM       & dPIRM               & dZIPIRM             \\\hline
			\multirow{3}{*}{0.1}        & 0.3 & 0.79(0.16) & 0.89(0.14)          & \textbf{0.91}(0.13) \\
			& 0.5 & 0.77(0.16) & \textbf{0.94}(0.10) & 0.93(0.10)          \\
			& 0.7 & 0.76(0.18) & 0.90(0.14)          & \textbf{0.94}(0.11) \\\hline
			\multirow{3}{*}{0.2}        & 0.3 & 0.77(0.16) & 0.87(0.15)          & \textbf{0.96}(0.07) \\
			& 0.5 & 0.76(0.16) & \textbf{0.94}(0.10) & 0.93(0.11)          \\
			& 0.7 & 0.76(0.16) & 0.92(0.12)          & \textbf{0.96}(0.08) \\\hline
			\multirow{3}{*}{0.3}        & 0.3 & 0.73(0.16) & \textbf{0.89}(0.14) & 0.88(0.13)          \\
			& 0.5 & 0.73(0.17) & 0.89(0.14)          & \textbf{0.90}(0.13) \\
			& 0.7 & 0.74(0.17) & 0.87(0.15)          & \textbf{0.90}(0.14)\\\hline
		\end{tabular}
		\label{tab1}
	\end{center}
\end{table}
\renewcommand{\arraystretch}{1.5}
\begin{table}[]
	\caption{Log-likelihood results}
	\begin{center}
		\begin{tabular}{cc|lll}\hline
			\multicolumn{1}{l}{$m$} & $s$ & PIRM        & dPIRM                & dZIPIRM              \\\hline
			\multirow{3}{*}{0.1}        & 0.3 & -2.28(0.38) & -2.04(0.28)          & \textbf{-2.01}(0.25) \\
			& 0.5 & -2.23(0.40) & \textbf{-2.02}(0.27) & \textbf{-2.02}(0.27) \\
			& 0.7 & -2.26(0.32) & -2.10(0.25)          & \textbf{-2.09}(0.25) \\\hline
			\multirow{3}{*}{0.2}        & 0.3 & -2.24(0.33) & -2.04(0.26)          & \textbf{-2.00}(0.25) \\
			& 0.5 & -2.20(0.33) & \textbf{-2.02}(0.23) & -2.04(0.25)          \\
			& 0.7 & -2.20(0.30) & -2.04(0.27)          & \textbf{-2.03}(0.29) \\\hline
			\multirow{3}{*}{0.3}        & 0.3 & -2.14(0.25) & -2.12(0.22)          & \textbf{-2.11}(0.22) \\
			& 0.5 & -2.22(0.25) & -2.09(0.27)          & \textbf{-2.08}(0.28) \\
			& 0.7 & -2.18(0.26) & \textbf{-2.02}(0.26) & -2.03(0.25)        \\\hline
		\end{tabular}
		\label{tab2}
	\end{center}
\end{table}
Tables \ref{tab1} and \ref{tab2} show the results for the RI and log-likelihood, respectively. $m$ represents the rate at which the object moved between clusters at each time step and $s$ represents the ratio of objects converted into zeros.

As shown in Table \ref{tab1}, the RIs of the dPIRM and dZIPIRM are higher than that of the PIRM in every setting. This finding means that models incorporating time structures such as the dPIRM and dZIPIRM can cluster data better than those that cluster data at each time step, the PIRM. When comparing the dPIRM and dZIPIRM, the dZIPIRM has a higher RI than the dPIRM in several settings. Furthermore, the dZIPIRM often yields a lower standard deviation than the dPIRM. This finding means that the dZIPIRM can perform stable clustering with accurate estimates even when many zero values exist in the data.

Table \ref{tab2} shows that the dPIRM and dZIPIRM yield higher log-likelihood values than the PIRM in any setting, just as with the RI. This means that we can better describe the assumed scenario by employing models incorporating time structures such as the dPIRM and dZIPIRM. When $m=0.1$, the dZIPIRM yields high log-likelihood values for every zero ratio. This finding shows that the dZIPIRM fits the data well when considering zero-inflated data for objects unlikely to move at each time step. Further, the dZIPIRM has higher log-likelihood values than the dPIRM for many patterns, like the RI, while the standard deviation is similar between the dPIRM and dZIPIRM, unlike the RI.

These results suggest that both the dPIRM and the dZIPIRM are superior to the PIRM in terms of RI and log-likelihood for this simulation setting. In addition, the dZIPIRM provides an accurate and stable clustering allocation as well as a good data fit compared with the dPIRM when time structures and zero-inflated data are assumed.
\section{Application}
In this section, we illustrate how the dPIRM and dZIPIRM may be applied to POS data to identify a group of products likely to be purchased simultaneously.
\subsection{Data and setting}
We analyze the POS data of a fashion brand in Japan, provided by Sothink Co., Ltd. These data provide information for two store types: shop-in-shop and franchise. Since fashion brand products are sensitive to the influence of the season in Japan, it is also important to consider marketing strategy changes depending on the season.

The data cover 78,807 customers and 12,937 products. Product categories  include short pants, small items, and short-sleeved T-shirts. We view product categories as objects in the model and count simultaneous purchases by product category.

We formulate a matrix showing the number of simultaneous purchases of 40 types of objects every month from January to June 2016 ($T=6$, $N=40$). That is, we create an $N\times N$ relational count data matrix for $T$ time steps. The zero-element ratio for the entire matrix is 0.52. We analyze the resulting dataset using the dPIRM and dZIPIRM. Since there is no prior hyperparameter information, we reuse the simulation approach.
\subsection{Results}
In this section, we first compare the log-likelihood values of the PIRM, dPIRM, and dZIPIRM to assess the fit of each model.
\begin{table}
	\caption{Log-likelihood results for the real data}
	\begin{center}
		\begin{tabular}{l|rrr}\hline
			& PIRM  & dPIRM & dZIPIRM        \\\hline
			Log-likelihood & -6.88 & -4.84 & \textbf{-3.84}\\\hline
		\end{tabular}
		\label{tb:table2}
	\end{center}
\end{table}

Table \ref{tb:table2} shows that the dZIPIRM has the highest log-likelihood, meaning that it provides the better fit to the data. Hence, assuming a ZIP and a time-varying structure improves the overall fit.
\begin{figure}[htbp]
	\begin{center}	
		\includegraphics[width=3.0in]{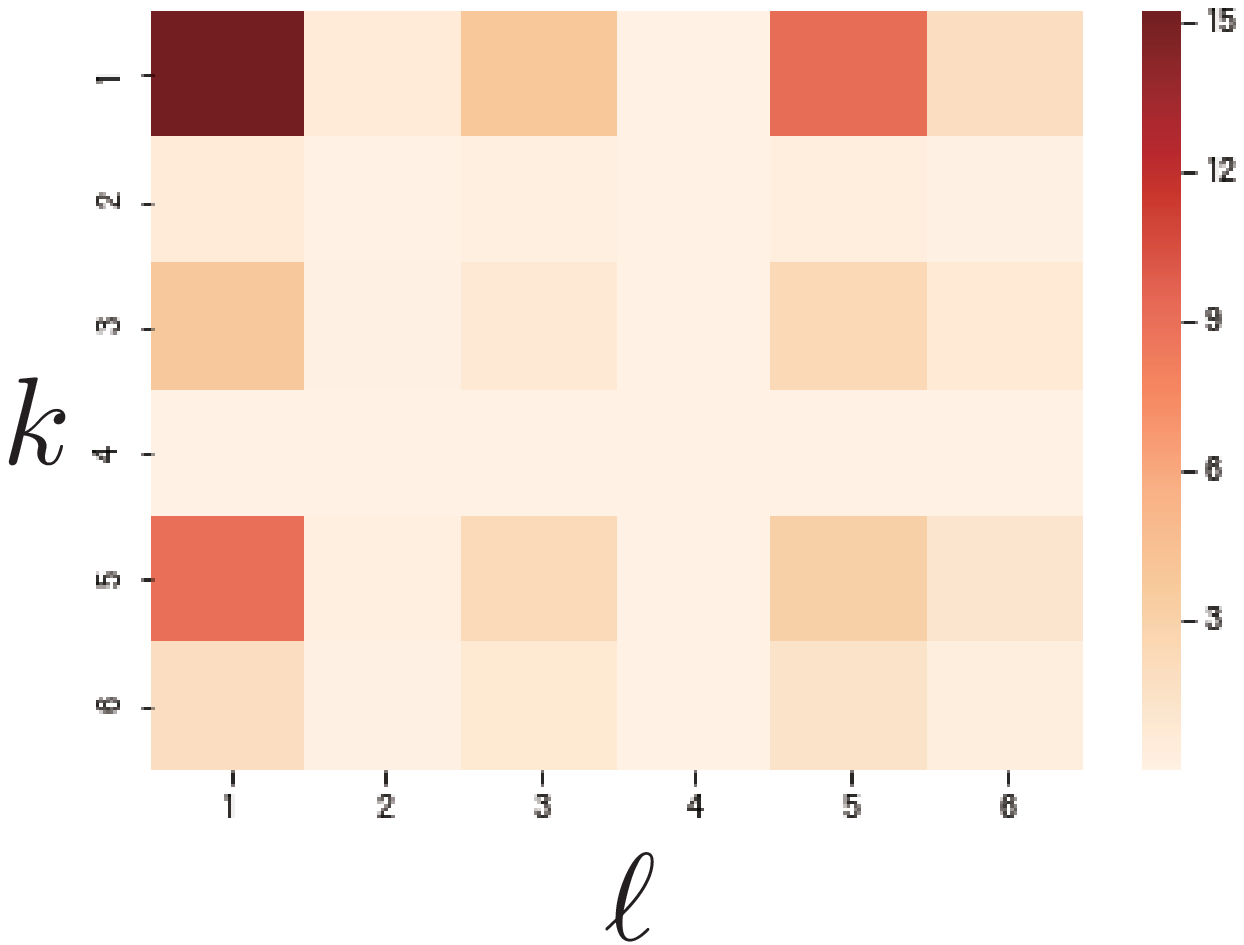}\\
		\caption{Heatmap of the estimated $\lambda_{k\ell}$ in the dPIRM}
		\label{fig:de1}
	\end{center}
\end{figure}
\begin{figure}[htbp]
	\begin{center}	
		\includegraphics[width=3.0in]{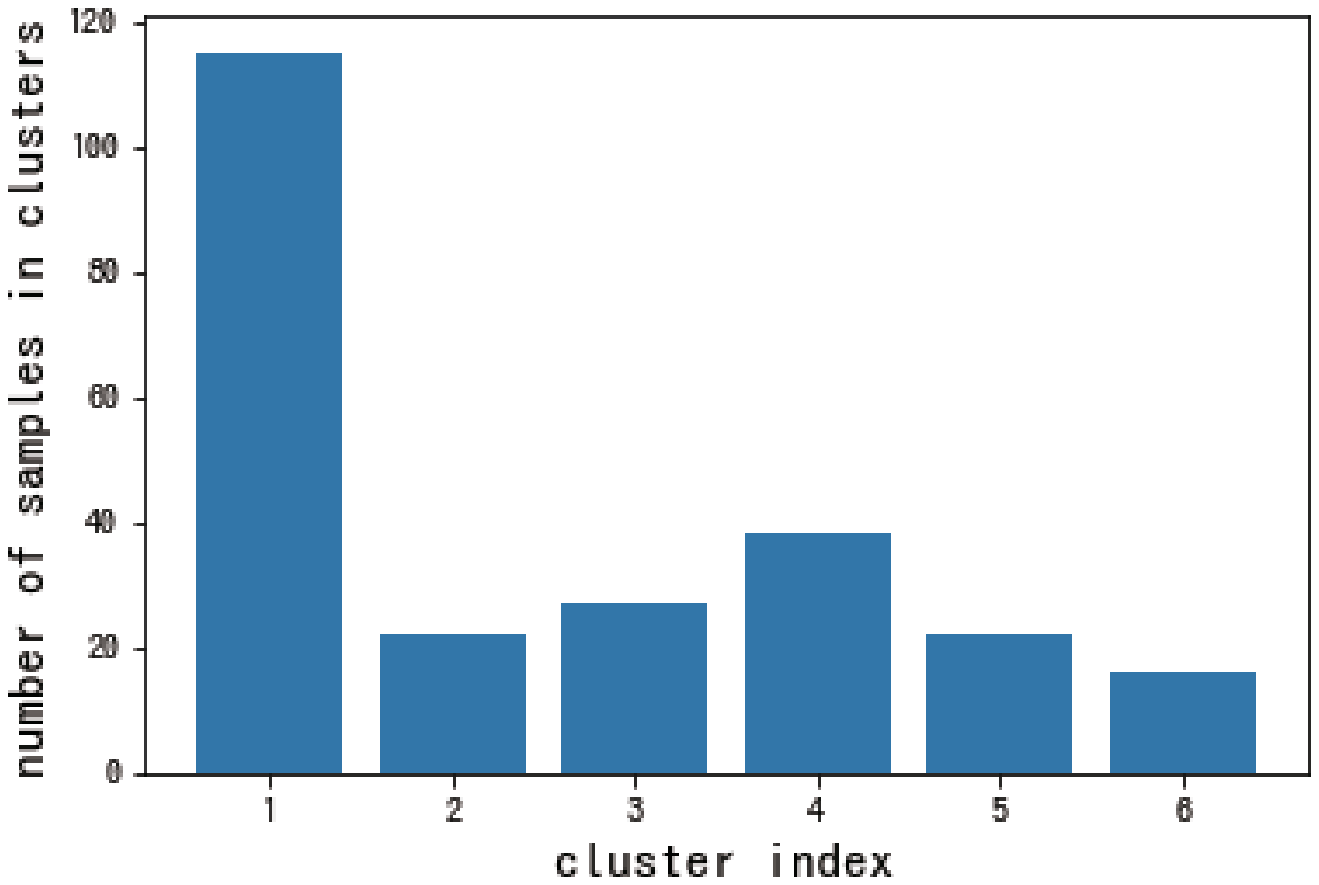}\\
		\caption{Total number of objects belonging to the clusters in the dPIRM}
		\label{fig:de2}
	\end{center}
\end{figure}
\begin{landscape}
	\renewcommand{\arraystretch}{2}
	\begin{table}[]
		\caption{Time-varying clustering assignments for selected objects in the dPIRM}
		\large
		\begin{center}
			\scalebox{0.85}{
				\begin{tabular}{lllllll}
					& \textbf{January}     & \textbf{February}    & \textbf{March}       & \textbf{April}       & \textbf{May}         & \textbf{June}        \\\cline{2-7} 
					\multicolumn{1}{c|}{\multirow{8}{*}{\textbf{cluster 1}}} & small item           & small item           & small item           & small item           & small item           &  \multicolumn{1}{l|}{small item}           \\
					\multicolumn{1}{c|}{}                                    & tie                  & tie                  & tie                  & tie                  & tie                  &  \multicolumn{1}{l|}{tie}                  \\
					\multicolumn{1}{c|}{}                                    & suit                 & suit                 & suit                 & suit                 & suit                 &  \multicolumn{1}{l|}{suit}                 \\
					\multicolumn{1}{c|}{}                                    & casual shirt         & casual shirt         & casual shirt         & casual shirt         & casual shirt         &  \multicolumn{1}{l|}{casual shirt}         \\
					\multicolumn{1}{c|}{}                                    & jeans                &                      & jeans                & jeans                & jeans                &  \multicolumn{1}{l|}{jeans}                \\
					\multicolumn{1}{c|}{}                                    &                      &                      &                      & slacks               & slacks               &  \multicolumn{1}{l|}{slacks}               \\
					\multicolumn{1}{c|}{}                                    &                      &                      &                      & short sleeve T-shirt & short sleeve T-shirt &  \multicolumn{1}{l|}{short sleeve T-shirt} \\
					\multicolumn{1}{c|}{}                                    &                      &                      &                      &                      &                      &  \multicolumn{1}{l|}{short pants}          \\\cline{2-7} 
					\multicolumn{1}{l|}{\multirow{2}{*}{\textbf{cluster 2}}}                     &                      &                      & short pants          & short pants          &                      &\multicolumn{1}{l|}{}                      \\ 
					\multicolumn{1}{l|}{}&                      &                      &                      & long sleeve T-shirt  &                      &    \multicolumn{1}{l|}{}                  \\\cline{2-7}
					\multicolumn{1}{l|}{\multirow{2}{*}{\textbf{cluster 3}}}                     & slacks               & slacks               &                      &                      &                      & \multicolumn{1}{l|}{}                     \\
					\multicolumn{1}{l|}{}& short sleeve T-shirt & short sleeve T-shirt &                      &                      &                      &  \multicolumn{1}{l|}{}                    \\\cline{2-7}
					\multicolumn{1}{l|}{\textbf{cluster 4}}                                      & short pants          & short pants          &                      &                      & long sleeve T-shirt  & \multicolumn{1}{l|}{long sleeve T-shirt}  \\\cline{2-7}
					\multicolumn{1}{l|}{\multirow{2}{*}{\textbf{cluster 5}}}                    &                      & jeans                & slacks               &                      & short pants          & \multicolumn{1}{l|}{}                     \\
					\multicolumn{1}{c|}{}&                      &                      & short sleeve T-shirt &                      &                      &  \multicolumn{1}{l|}{}                    \\\cline{2-7}
					\multicolumn{1}{l|}{\textbf{cluster 6}}                                       & long sleeve T-shirt  & long sleeve T-shirt  & long sleeve T-shirt  &                      &                      & \multicolumn{1}{l|}{}                    \\\cline{2-7}
				\end{tabular}}
				\label{fig:idou}
			\end{center}
		\end{table}
	\end{landscape}
	
	Next, we show the results for the dPIRM. The number of clusters in the dPIRM is estimated to be six. Figure \ref{fig:de1} shows the estimated $\lambda_{k\ell}$ corresponding to each cluster combination, while Figure \ref{fig:de2} shows the total number of objects in each cluster for the six-month sample period. 
	Table \ref{fig:idou} shows the objects belonging to the corresponding clusters.
	Since the parameter $\lambda_{k,\ell}$ represents the strength of the relationship, if the value of the estimated $\lambda_{k\ell}$ is large, the objects in Cluster $k$ and Cluster $\ell$ are more likely to be purchased simultaneously.

	Figure \ref{fig:de1} shows that the objects in Cluster 1 are more likely to be purchased with those in Cluster 1 simultaneously (i.e., $\hat{\lambda}_{11}$ is high). Further, the objects in Cluster 3 are more likely to be purchased simultaneously with those in Cluster 1 and Cluster 5 (i.e., $\hat{\lambda}_{31}$ and $\hat{\lambda}_{35}$ are high), while they are unlikely to be purchased with other items in Cluster 3 (i.e., $\hat{\lambda}_{33}$ is low). The items in Cluster 5 are more likely to be bought with items in Cluster 1, followed in order of those in Clusters 5 and 3 (i.e., $\hat{\lambda}_{51}>\hat{\lambda}_{55}>\hat{\lambda}_{53}$).
	
	From the number of objects belonging to each cluster in Figure \ref{fig:de2}, we see that the number in Cluster 1 is much larger than that for the other clusters. Therefore, the objects in Cluster 1 are more likely to be purchased simultaneously with several other types of objects. As shown in Table \ref{fig:idou}, there are two types of objects in Cluster 1: objects that remain in Cluster 1 for a long period and items that move from or to Cluster 1 depending on the season.

	Examples of objects that remain in Cluster 1 for a long period include small items such as handkerchiefs and necklaces, which may be interpreted as objects likely to be purchased along with other items throughout the year. We may also interpret items such as ties as being more likely to be bought with a suit throughout the year.
	
	Meanwhile, items that move from or to Cluster 1 depending on the season, such as slacks and short-sleeved T-shirts, belong to Clusters 3 and 5 at the beginning of the year, but move to Cluster 1 from April onwards. This date is the turning point of the season in Japan, when spring starts and temperatures rise. This means that slacks and short-sleeved T-shirts are not purchased much in the winter, whereas in the spring they are more likely to be bought along with other objects.
	\begin{figure}[htbp]
		\begin{center}	
			\includegraphics[width=3.0in]{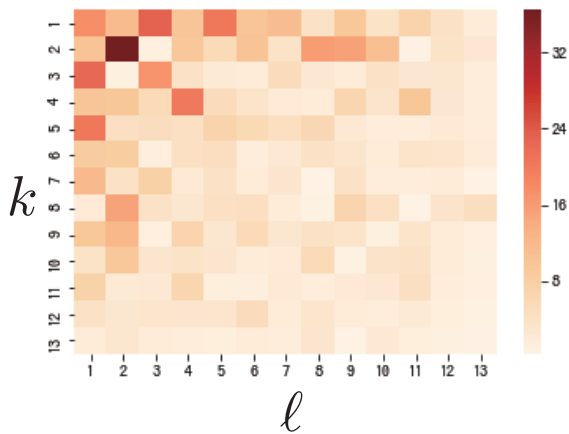}\\
			\caption{Heatmap of the estimated $\lambda_{k\ell}$ in the dZIPIRM}
			\label{fig:dede1}
		\end{center}
	\end{figure}
	\begin{figure}[htbp]
		\begin{center}	
			\includegraphics[width=3.0in]{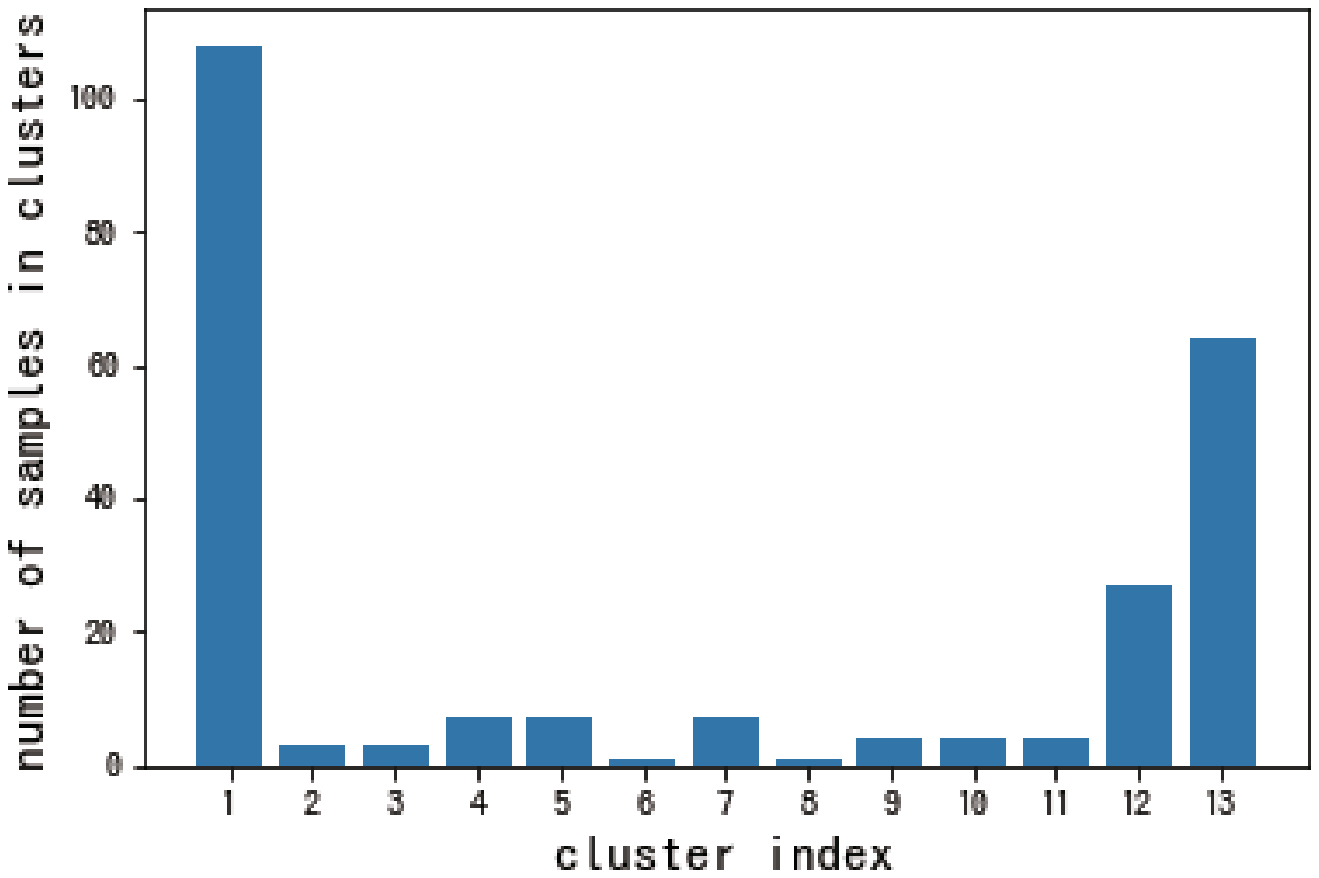}\\
			\caption{Total number of objects belonging to the clusters in the dZIPIRM}
			\label{fig:dede2}
		\end{center}
	\end{figure}
	\begin{landscape}
		\renewcommand{\arraystretch}{1.8}
		\begin{table}[]
			\caption{Time-varying clustering assignments for selected objects in the dZIPIRM}
			\large
			\begin{center}
				\scalebox{0.85}{
					\begin{tabular}{lllllll}
						& \textbf{January}     & \textbf{February}    & \textbf{March}       & \textbf{April}       & \textbf{May}         & \textbf{June}        \\\cline{2-7} 
						\multicolumn{1}{c|}{\multirow{8}{*}{\textbf{cluster 1}}} & small item           & small item           & small item           & small item           & small item           & \multicolumn{1}{l|}{small item}           \\
						\multicolumn{1}{c|}{}                                    & tie                  & tie                  & tie                  & tie                  & tie                  & \multicolumn{1}{l|}{tie}                  \\
						\multicolumn{1}{c|}{}                                    & suit                 & suit                 & suit                 & suit                 & suit                 & \multicolumn{1}{l|}{suit}                 \\
						\multicolumn{1}{c|}{}                                    &                      &                      & casual shirt         & casual shirt         & casual shirt         & \multicolumn{1}{l|}{casual shirt}         \\
						\multicolumn{1}{c|}{}                                    &                      &                      & jeans                & jeans                & jeans                & \multicolumn{1}{l|}{jeans}                \\
						\multicolumn{1}{c|}{}                                    &                      &                      &                      & slacks               & slacks               & \multicolumn{1}{l|}{slacks}               \\
						\multicolumn{1}{c|}{}                                    &                      &                      &                      & short sleeve T-shirt & short sleeve T-shirt & \multicolumn{1}{l|}{short sleeve T-shirt} \\
						\multicolumn{1}{c|}{}                                    &                      &                      &                      &                      & short pants          & \multicolumn{1}{l|}{short pants}          \\\cline{2-7}
						\multicolumn{1}{c|}{\multirow{2}{*}{\textbf{cluster 2}}}                     & casual shirt         &                      &                      & short sleeve T-shirt & short sleeve T-shirt & \multicolumn{1}{l|}{short sleeve T-shirt} \\
						\multicolumn{1}{l|}{}& jeans                &                      &                      &                      &                      & \multicolumn{1}{l|}{short pants}          \\\cline{2-7}
						\multicolumn{1}{c|}{\multirow{2}{*}{\textbf{cluster 4}}}                     &                      & casual shirt         &                      &                      &                      &\multicolumn{1}{l|}{}                      \\
						\multicolumn{1}{l|}{}&                      & jeans                &                      &                      &                      & \multicolumn{1}{l|}{}                     \\\cline{2-7}
						\multicolumn{1}{l|}{\textbf{cluster 7}}                                      &                      &                      & slacks               &                      &                      &\multicolumn{1}{l|}{}                      \\\cline{2-7}
						\multicolumn{1}{l|}{\textbf{cluster 9}}                                      &                      & slacks               &                      &                      &                      & \multicolumn{1}{l|}{}                     \\\cline{2-7}
						\multicolumn{1}{l|}{\textbf{cluster 10}}                                     & short sleeve T-shirt & short sleeve T-shirt &                      &                      &                      & \multicolumn{1}{l|}{}                     \\\cline{2-7}
						\multicolumn{1}{l|}{\textbf{cluster 11}}                                     &                      &                      & short sleeve T-shirt &                      & short pants          &  \multicolumn{1}{l|}{}                    \\\cline{2-7}
						\multicolumn{1}{l|}{\textbf{cluster 12}}                                     & slacks               &                      & long sleeve T-shirt  & short pants          &                      & \multicolumn{1}{l|}{}                     \\\cline{2-7}
						\multicolumn{1}{c|}{\multirow{2}{*}{\textbf{cluster 13}}}                    & short pants          & short pants          & short pants          & long sleeve T-shirt  & long sleeve T-shirt  & \multicolumn{1}{l|}{long sleeve T-shirt}  \\
						\multicolumn{1}{l|}{}& long sleeve T-shirt  & long sleeve T-shirt  &                      &                      &                      &   \multicolumn{1}{l|}{}\\\cline{2-7}                  
					\end{tabular}}
					\label{fig:idou2}
				\end{center}
			\end{table}
		\end{landscape}
		
		We now show the results for the dZIPIRM.  The estimated number of clusters is 13. Figures \ref{fig:dede1}, \ref{fig:dede2} and Table \ref{fig:idou2} for the dZIPIRM follow Figures \ref{fig:de1}, \ref{fig:de2} and Table \ref{fig:idou} for the dPIRM.  Figure \ref{fig:dede1} shows that the number of clusters in the dZIPIRM is larger than that in the dPIRM because of the increase in cluster expressiveness, which in turn is a consequence of the additional flexibility provided by the assumption that there are many zeros.

		Beyond the estimated number of clusters, the results for the dZIPIRM are similar as those for the dPIRM. For example, Cluster 1 of the dZIPIRM is similar to Cluster 1 of the dPIRM (e.g., always including small items, ties, and suits), while some objects move from cluster to cluster depending on the season. However, using the dZIPIRM, we may interpret the result in more detail. For example, the dPIRM could not distinguish between the behavior of slacks and short-sleeved T-shirts, whereas the dZIPIRM can interpret the difference between these objects. Specifically, in January, the category ``slacks'' belongs to Cluster 12, while the category ``short-sleeved T-shirts'' belongs to Cluster 10. Products in Cluster 12 are unlikely to be purchased with objects in any cluster. Meanwhile, objects in Cluster 10 are likely to be purchased with objects in Cluster 2 (e.g., casual shirts and jeans). Thus, by assuming a zero-inflated setting using the dZIPIRM, we obtain a result that is easier to interpret.

		From the above analysis, we may conclude that since the dPIRM yields fewer clusters, only rough trends may be modeled. However, as the dZIPIRM yields more clusters, a more detailed interpretation of the clusters may be obtained. As Table \ref{tb:table2} shows, the dZIPIRM fits the data better than the dPIRM. Despite this, as is typically the case with empirical data, it is difficult to objectively validate which method should be preferred. Nevertheless, this empirical application shows that both the dPIRM and the dZIPIRM allow us to see how customer buying behavior changes with the seasons. In addition, by using the dZIPIRM, we obtain a clustering result that provides a more detailed interpretation of the clusters when the data contain many zeros.
		\section{Conclusion and Discussion}
		We proposed two bi-clustering methods for count data considering changes over time. The first method, the dPIRM, enabled the bi-clustering of time-varying count data by using the time structure and distribution information for these data. The second method, the dZIPIRM, extended the dPIRM to express zero-inflated real data. To derive the posterior distribution and clarify the properties of the parameters, zero-inflated data were formulated by using latent variables.

		We derived the explicit posterior distributions of the parameters and outlined their properties under the estimation framework using a full conditional distribution. Specifically, by comparing the expected value and variance of $\lambda$ for both the dPIRM and the dZIPIRM, we showed that how the specific model assumptions affect the estimation result. We also explained the theoretical relationship between the two models. Although this theorem is specifically derived to compare our two proposed model, dPIRM and dZIPIRM, this could be also applied for other ZIP-related model.
		
		By incorporating a time structure, we further showed that both the dPIRM and the dZIPIRM outperform the PIRM in terms of the RI and log-likelihood values in a simulation study. We also observed that the dZIPIRM yields more stable and better clustering allocations as well as a better data fit than the dPIRM.

		For the empirical data, the dZIPIRM, which assumes the properties of the ZIP and a time-varying structure, resulted in the best data fit. Moreover, the clustering result of the dZIPIRM is similar to that of the dPIRM. However, using the dZIPIRM allows us to interpret the cluster structure in more detail, as this model relies on more flexible assumptions than the dPIRM.
		
		We consider two future works. At first, although the row and column objects in the dPIRM and dZIPIRM may differ, for comprehensibility, we analyzed the data using the same object for the rows and columns.
		Therefore, future work will focus on real data in situations where the objects of the rows and columns are different.

		Future research will also study how the choice of hyperparameters affects the estimation results in a simulation setting (i.e., how the estimation result changes depending on the choice of hyperparameters). In Section 3.2, we referred to the relationship between the hyperparameters and parameters, and we believe further work on this area is needed to provide users with better guidance about hyperparameter choice.
		\section*{Appendix}
		In this Appendix, we show proofs for Equation (\ref{rtox}) and Theorem 1.
		First, Equation (\ref{rtox}) can be derived as follows.
		\begin{proof}
			\begin{align*}
				\sum_{t,i,j\in C_{k\ell}}r_{tij}x_{tij}&=\sum_{t,i,j\in C_{k\ell};x_{tij}=0}r_{tij}x_{tij}+\sum_{t,i,j\in C_{k\ell};x_{tij}>0}r_{tij}x_{tij}\notag\\
				&=\sum_{t,i,j\in C_{k\ell};x_{tij}>0}r_{tij}x_{tij}\notag\\
				&=\sum_{t,i,j\in C_{k\ell};x_{tij}>0}x_{tij}\quad (x_{tij}>0\Longrightarrow r_{tij}=1)\\
				&=\sum_{t,i,j\in C_{k\ell};x_{tij}=0}x_{tij}+\sum_{t,i,j\in C_{k\ell};x_{tij}>0}x_{tij}\notag\\
				&=\sum_{t,i,j\in C_{k\ell}}x_{tij}
			\end{align*}
		\end{proof}

		Next, proof of Theorem 1 is derived as follows.
		\begin{proof}
			\begin{align}
				\label{tenkai}
				\sum_{t,i,j\in C_{k\ell}}r_{tij}=\sum_{t,i,j}I\{t,i,j\in C_{k\ell}\}-\sum_{t,i,j\in C_{k\ell}}(1-r_{tij})
			\end{align}
			Here, we derive Eq. (\ref{the1})
			\begin{align*}
				E\left[\lambda_{k\ell}^{(dP)}\right]&=\frac{a+\sum_{t,i,j\in C_{k\ell}}x_{tij}}{b+\sum_{t,i,j}I\{t,i,j\in C_{k\ell}\}}\\
				&=\frac{a+\sum_{t,i,j\in C_{k\ell}}x_{tij}}{b+\sum_{t,i,j\in C_{k\ell}}r_{tij}}\\
				&\quad+\left(\frac{a+\sum_{t,i,j\in C_{k\ell}}x_{tij}}{b+\sum_{t,i,j}I\{t,i,j\in C_{k\ell}\}}-\frac{a+\sum_{t,i,j\in C_{k\ell}}x_{tij}}{b+\sum_{t,i,j\in C_{k\ell}}r_{tij}}\right)\\
				&=E\left[\lambda_{k\ell}^{(dZIP)}\right]-\Biggl[\left(a+\sum_{t,i,j\in C_{k\ell}}
				x_{tij}\right)\\
				&\quad\times \frac{\sum_{t,i,j}I\{t,i,j\in C_{k\ell}\}-\sum_{t,i,j\in C_{k\ell}}r_{tij}}{\left(b+\sum_{t,i,j}I\{t,i,j\in C_{k\ell}\}\right)\left(b+\sum_{t,i,j\in C_{k\ell}}r_{tij}\right)}\Biggr]\\
				&=E\left[\lambda_{k\ell}^{(dZIP)}\right]-\left(\frac{\sum_{t,i,j\in C_{k\ell}}(1-r_{tij})}{b+\sum_{t,i,j}I\{t,i,j\in C_{k\ell}\}}\right)E\left[\lambda_{k\ell}^{(dZIP)}\right]\,&(by \,Eq. (\ref{tenkai}))\\
				&=\left(1-\frac{\sum_{t,i,j\in C_{k\ell}}(1-r_{tij})}{b+\sum_{t,i,j}I\{t,i,j\in C_{k\ell}\}}\right)E\left[\lambda_{k\ell}^{(dZIP)}\right]\\
				&=\left(\frac{b+\sum_{t,i,j\in C_{k\ell}}r_{tij}}{b+\sum_{t,i,j}I\{t,i,j\in C_{k\ell}\}}\right)E\left[\lambda_{k\ell}^{(dZIP)}\right]
			\end{align*}
			Here, we derive Eq. (\ref{the2})
			\begin{align*}
				V\left[\lambda_{k\ell}^{(dP)}\right]&=\frac{a+\sum_{t,i,j\in C_{k\ell}}x_{tij}}{\left(b+\sum_{t,i,j}I\{t,i,j\in C_{k\ell}\}\right)^2}\\
				&=\frac{a+\sum_{t,i,j\in C_{k\ell}}x_{tij}}{\left(b+\sum_{t,i,j\in C_{k\ell}}r_{tij}\right)^2}\\
				&\quad+\left(\frac{a+\sum_{t,i,j\in C_{k\ell}}x_{tij}}{\left(b+\sum_{t,i,j}I\{t,i,j\in C_{k\ell}\}\right)^2}-\frac{a+\sum_{t,i,j\in C_{k\ell}}x_{tij}}{\left(b+\sum_{t,i,j\in C_{k\ell}}r_{tij}\right)^2}\right)\\
				&=V\left[\lambda_{k\ell}^{(dZIP)}\right]-\Biggl[\left(a+\sum_{t,i,j\in C_{k\ell}}x_{tij}\right)\\
				&\quad\times \frac{\left(b+\sum_{t,i,j}I\{t,i,j\in C_{k\ell}\}\right)^2-\left(b+\sum_{t,i,j\in C_{k\ell}}r_{tij}\right)^2}{\left(b+\sum_{t,i,j}I\{t,i,j\in C_{k\ell}\}\right)^2\left(b+\sum_{t,i,j\in C_{k\ell}}r_{tij}\right)^2}\Biggr]\\
				&=V\left[\lambda_{k\ell}^{(dZIP)}\right]\\
				&\quad -\Biggl[\frac{\left(2b+\sum_{t,i,j}I\{t,i,j\in C_{k\ell}\}+\sum_{t,i,j\in C_{k\ell}}r_{tij}\right)}{\left(b+\sum_{t,i,j}I\{t,i,j\in C_{k\ell}\}\right)^2}\\
				&\quad\quad\quad\times \sum_{t,i,j\in C_{k\ell}}(1-r_{tij})V\left[\lambda_{k\ell}^{(dZIP)}\right]\Biggr]\,&(by \,Eq. (\ref{tenkai}))\\
				&=\left(\frac{b+\sum_{t,i,j\in C_{k\ell}}r_{tij}}{b+\sum_{t,i,j}I\{t,i,j\in C_{k\ell}\}}\right)^2V\left[\lambda_{k\ell}^{(dZIP)}\right]
			\end{align*}
			\begin{align*}
				0&\leq\sum_{t,i,j\in C_{k\ell}}r_{tij}\leq\sum_{t,i,j}I\{t,i,j\in C_{k\ell}\}\quad(Eq. (\ref{tenkai}))\\
				0&\leq\frac{b+\sum_{t,i,j\in C_{k\ell}}r_{tij}}{b+\sum_{t,i,j}I\{t,i,j\in C_{k\ell}\}}\leq 1
			\end{align*}
		\end{proof}


\begin{thebibliography}{30}
\bibitem{bayeszip}
Angers, J. F., Biswas, A., 2003. A Bayesian analysis of zero-inflated generalized Poisson model.
{\it Computational Statistics and Data Analysis},
			{\bf 42}, 37--46. 
			
			\bibitem{zipdif}
			Bhattacharya, A., Clarke, B. S., Datta, G., 2008. A Bayesian test for excess zeros in a zero-inflated power series distribution.
			{\it IMS Collections},
			{\bf 1}, 89--104. 
			
			\bibitem{hmm2}
			Fox, E. B., Sudderth, E. B., Jordan, M. I., Willsky, A. S., 2008.
			An HDP-HMM for systems with state persistence.
			 In {\it Proceedings of the 25th International Conference on Machine Learning},
			312--319. 
			
			\bibitem{zi2}
			Fox, J. P., 2013.
			Multivariate zero-inflated modeling with latent predictors: modeling feedback behavior.
			{\it Computational Statistics and Data Analysis},
			{\bf 68}, 361--374. 
			
			\bibitem{DMMB}
			Fu, W., Song, L., Xing, E. P., 2009.
			Dynamic mixed membership blockmodel for evolving networks.
			In {\it Proceedings of the 26th Annual International Conference on Machine Learning},
			329--336. 
			
			\bibitem{biclu2}
			Hartigan, J. A., 1972.
			Direct clustering of a data matrix.
			{\it Journal of the American Statistical Association},
			{\bf 67}(337), 123--129. 
			
			\bibitem{dIRM}
			Ishiguro, K., Iwata, T., Ueda, N., Tenenbaum, J. B., 2010.
			Dynamic infinite relational model for time-varying relational data analysis.
			In {\it Proceedings of Advances in Neural Information Processing Systems},
			919--927. 
			
			\bibitem{SIRM}
			Ishiguro, K., Ueda, N., Sawada, H., 2012.
			Subset infinite relational models.
			In {\it Proceedings of Artificial Intelligence and Statistics},
			547--555. 
			
			\bibitem{IRM}
			Kemp, C., Tenenbaum, J. B., Griffiths, T. L., Yamada, T., Ueda, N., 2006.
			Learning systems of concepts with an infinite relational model.
			In {\it Proceedings of the 21st National Conference on Artificial Intelligence},
			381--388.
			
			\bibitem{ZIP}
			Lambert, D., 1992.
			Zero-inflated Poisson regression, with an application to defects in manufacturing.
			{\it Technometrics},
			{\bf 34}(1), 1--14.
			
			\bibitem{zi1}
			Lee, K., Joo, Y., Song, J. J., Harper, D. W., 2011.
			Analysis of zero-inflated clustered count data: a marginalized model approach.
			{\it Computational Statistics and Data Analysis},
			{\bf 55}, 824--837.
			
			
			\bibitem{bayzi}
			Liu, Y., Tian, G. L., 2015.
			Type I multivariate zero-inflated Poisson distribution with applications.
			{\it Computational Statistics and Data Analysis},
			{\bf 83}, 200--222.
			
			\bibitem{RI}
			Rand, W. M., 1971.
			Objective criteria for the evaluation of clustering methods.
			{\it Journal of the American Statistical Association},
			{\bf 66}, 846--850.
			
			\bibitem{biclu1}
			Boris, M., 1996. {\it Mathematical Classification and Clustering}.
			Kluwer academic publishers,
			Norwell, USA.
			
			\bibitem{SBM}
			Nowicki, K., Snijders, T. A. B., 2001.
			Estimation and prediction for stochastic blockstructures.
			{\it Journal of the American Statistical Association},
			{\bf 96}(455), 1077--1087.
			
			\bibitem{ZIPhmm}
			Olteanu, M., Ridgway, J., 2012.
			Hidden markov models for time series of counts with excess zeros.
			In {\it Proceedings of 20th European Symposium on Artificial Neural Networks},
			133--138.
			
			\bibitem{stick}
			Sethuraman, J., 1994.
			A constructive definition of Dirichlet piors.
			{\it Statistica Sinica},
			{\bf 4}, 639--650.
			
			\bibitem{hmm1}
			Teh, Y. W., Jordan, M. I., Beal, M. J., Blei, D. M., 2007.
			Hierarchical Dirichlet processes.
			{\it Journal of the American Statistical Association},
			{\bf 101}(476), 1566--1581.
			
			\bibitem{beam}
			Van Gael, J., Saatci, Y., Teh, Y. W., Ghahramani, Z., 2008.
			Beam sampling for the infinite hidden Markov model.
			In {\it Proceedings of the 25th International Conference on Machine Learning},
			1088--1095.
			
			\bibitem{JAT}
			Wang, E., Liu, D., Sliva, J., Carin, L., Dunson, D. B., 2010.
			Joint analysis of time-evolving binary matrices and associated documents.
			In {\it Proceedings of Advances in Neural Information Processing Systems},
			2370--2378.
			
			\bibitem{PIRM}
			Wu, X., Li, H., 2017.
			Topic mover's distance based document classification.
			{\it 2017 IEEE 17th International Conference on Communication Technology (ICCT)},
			1998--2002.
			
			\bibitem{BAFC}
			Yang, T., Chi, Y., Zhu, S., Gong, Y., Jin, R., 2009.
			A Bayesian approach toward finding communities and their evolutions in dynamic social networks.
			In {\it Proceedings of the 2009 SIAM International Conference on Data Mining},
			990--1001.
\end{thebibliography}


\end{document}